\documentclass[%
reprint,
superscriptaddress,
amsmath,amssymb,
aps,
twocolumn
]{revtex4-2}

\usepackage{graphicx}
\usepackage{amssymb}
\usepackage{longtable}
\usepackage{rotating}
\usepackage{array}
\usepackage{multirow}
\usepackage{makecell}
\usepackage{booktabs}
\usepackage{xcolor}
\usepackage{ulem}
\usepackage{colortbl}
\usepackage{amsmath,esint}
\usepackage[flushleft]{threeparttable}
\usepackage{hyperref}
\usepackage{tikz}
\usepackage{float}

\setcitestyle{super}
\usepackage{dcolumn}
\usepackage{bm}
\usepackage{physics}
\usepackage{hyperref}
          
\newcommand*{\citen}[1]{%
  \begingroup
    \romannumeral-`\x 
    \setcitestyle{numbers}%
    \cite{#1}%
  \endgroup   
}

\begin{document}

\preprint{APS/123-QED}

\title{Photoelectron chiral dichroism induced by lasers without helicity via chiral hole wave-packets} 

\author{Gal Bouskila}\email{galboski@campus.technion.ac.il}
\affiliation{Schulich Faculty of Chemistry, Technion – Israel Institute of Technology, 32000 Haifa, Israel}
\author{Avner Fleischer}
\affiliation{
 Raymond and Beverly Sackler Faculty of Exact Sciences, School of Chemistry, Tel Aviv University, Tel-Aviv, Israel}
\author{Ofer Neufeld}\email{ofern@technion.ac.il}
\affiliation{Schulich Faculty of Chemistry, Technion – Israel Institute of Technology, 32000 Haifa, Israel}

\date{\today}

\begin{abstract}
Photoelectron circular dichroism (PECD) is a method where randomly oriented chiral molecules are photoionized due to irradiation by circularly-polarized lasers, yielding large chiral signals in the photoelectron momentum distribution. Recently, PECD was explored with polarization-tailored light such as bi-chromatic and non-collinear drivers, which still produces significant chiral signals. Yet, all known PECD configurations to date exhibit non-zero time-local chirality. That is, they are driven by an intrinsically helical light source. Nonetheless, 'chiral' light can also be non-helical if its chirality manifests on longer timescales (e.g. an optical centrifuge). It remains unknown whether PECD can arise from non-helical coherent light. Here we predict that PECD indeed emerges from non-helical light by employing a train of linearly-polarized intense laser pulses with a rotating polarization axis, which are phase-coherent and time-delayed. We find strong PECD in the model chiral molecule CBrClFH under a wide parameter regime that can be optimized up to $\sim$8\% by tuning delays between pulses, suggesting quantum interference. We directly show that the physical mechanism for this type of PECD differs from the standard case, relying on a chiral hole attosecond wave-packet evolving in the molecule, induced by the first linear pulse. Our work shows that multiple mechanisms can give rise to PECD on longer timescales and provides a novel approach for ultrafast chirality spectroscopy and coherent chiral wave-packet manipulation.
\end{abstract}

\maketitle

Chirality describes the asymmetry of objects that cannot be superimposed onto their mirror image~\cite{kelvin1894molecular}. It arises naturally throughout the macroscopic and microscopic world and plays a key role in materials science~\cite{Hosur2010,Wang2014}, chemistry~\cite{Meyer-Ilse2013}, and biology~\cite{Reddy2004,Guerrero-Martinez2011}. In chemistry and nanoscience, molecules of opposite handedness, denoted enantiomers, have identical chemical and physical properties, but differ in their interactions with other chiral objects such as chiral molecules or circularly-polarized light (CPL). This property is crucial for a variety of applications from drug design~\cite{mcbride1961thalidomide,h2011significance} to magnetism~\cite{casher1974chiral,yang2021chiral,li2016chiral}.

Traditionally, interactions with CPL have served as the primary method for detecting chirality~\cite{polavarapu2016chiroptical}. For example, chirality is detected through polarization rotation or circular dichroism (CD) in absorption~\cite{barron2009molecular}. However, these techniques are inherently limited by weak chiral signals, often on the order of ~0.001–0.1\%, because they rely on magnetic dipolar interactions~\cite{berova2000circular}. As a result, conventional CD methods face challenges in probing a wide range of chiral systems~\cite{ayuso2022ultrafast,baykusheva2019real,svoboda2022femtosecond,beaulieu2017attosecond,cireasa2015probing,quack2000influence,erez2023simultaneous} such as large bio-molecules with multiple chiral centers~\cite{neufeld2022detecting}.

Accordingly, in recent decades a variety of techniques have been developed to overcome these limitations\cite{Habibovic2024,Fischer2000,Perez2017,Ayuso2019,Herwig2013,neufeld2025light}. This Letter focuses on photoelectron CD (PECD)~\cite{Ritchie1976}, which has demonstrated robust gas-phase chiral signals up to ~30\%\cite{Bowering2001,Janssen2014} and can obtain ultrafast temporal resolution\cite{Comby2016,Facciala2023}. In PECD, CPL irradiates randomly oriented chiral media, producing photoelectrons that are preferentially emitted forward or backward relative to light’s propagation direction. This dichroism arises from an inherent chiral light-matter interaction of the photoelectron with the molecular chiral potential, and with the helical laser beam that is photoionizing it as it travels to the detector. By comparing normalized photoelectron spectrum (PES) for opposite CPL helicities or molecular handedness (R/S), the chiral signal is obtained as (Eq.~\ref{eq:pecd}): 

\begin{equation}
    PECD(\textbf{k}) =  2\cdot\frac{PES_{R}(\textbf{k})-PES_{S}(\textbf{k})}{PES_{R}(\textbf{k})+PES_{S}(\textbf{k})}
\label{eq:pecd}
\end{equation}
where $\textbf{k}$ is the photoelectron momentum. 

Recently, PECD has also been explored in configurations involving tailored bi-chromatic light fields~\cite{demekhin2018photoelectron,rozen2019controlling,hofmann2024subcycle, neufeld2021strong, ayuso2019synthetic}. Here, multiple carrier waves are combined to shape the optical field’s time-dependent polarization and instantaneous chirality~\cite{neufeld2018optical}, which still produces chiral signals on the order of a few percent~\cite{rozen2019controlling}. In these cases as well, the driving laser is helical. Mathematically, this helicity can be described using Optical chirality (OC)~\cite{tang2010optical,lipkin1964existence}, which is commonly used to quantify the local and instantaneous chirality of electromagnetic (EM) waves and is proportional to the helicity of the light~\cite{neufeld2018optical}. All PECD configurations measured to date employed helical driving lasers with OC$\neq$0. Nevertheless, chiral light can be produced even with OC=0 if the chiral structure is sustained on longer timescales, exhibiting non-instantaneous OC~\cite{neufeld2018optical}. For example, an optical centrifuge~\cite{milner2019controlled} exhibits a rotating linear polarization over time; its OC vanishes since it is linearly-polarized at any point in time, but is still chiral on longer timescales due to the gradual rotation of the polarization axis. PECD arising from a pair of linearly-polarized pulses, together forming a non-helical light field, was analytically briefly discussed in Ref.~\citen{ordonez2018generalized}, but has not been demonstrated or further explored. Additionally, the pulses considered in Ref.~\citen{ordonez2018generalized} were not phase-coherent, making it unclear under what realistic conditions PECD could manifest in such cases. From a fundamental standpoint, it remains an open question whether non-helical light (i.e. light with OC=0), can generate PECD. If such a response is possible, it is also unclear how the non-instantaneous optical chirality connects to observable signals, and what physical mechanisms might be responsible for generating them.

\begin{figure}[H]
    \centering
    \includegraphics[width=1.0\linewidth]{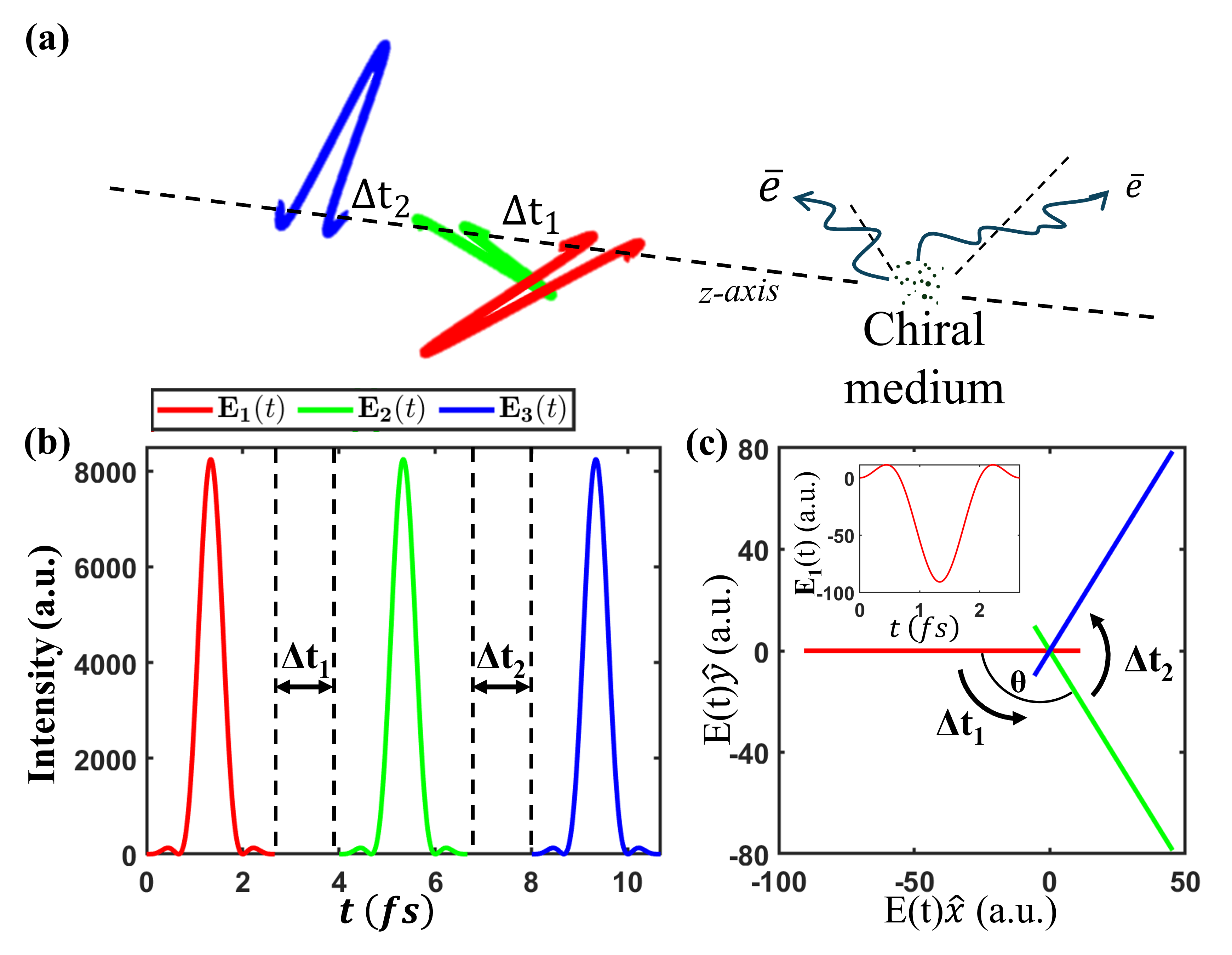}
    \caption{{\bf (a)  Illustration of laser pulse train set-up and emerging PECD. (b) Laser instantaneous intensity for the pulse train set-up. (c) Lissajous curve for pulse train. Inset shows the temporal evolution of single pulse out of the train.} Here each pulse has a peak intensity of $5\cdot10^{13} [W/cm^2]$, the angle between pulse polarizations is $\theta = 120^o$, $\omega$ corresponds to a wavelength of $800nm$, the inter-pulse delay is $T/2$ for both pulses with $T$ the optical cycle, and in each pulse CEP = $\pi$.} \label{fig:laser}
\end{figure}

Here we predict that PECD indeed emerges from a zero-OC light. We demonstrate this with \textit{ab-initio} calculations of PECD driven in bromochlorofluoromethane (CBrClFH), the simplest stable chiral molecule, driven by a train of linear-polarized intense laser pulses with a linear polarization axis that rotates from pulse to pulse (see Fig.~\ref{fig:laser}). By tuning the relative delays between pulses in the train, we obtain strong PECD signals across a wide range of conditions, reaching up to $\sim$8\% (with a dynamical range down to $\sim$2\%), consistent with an attosecond interference interpretation. By varying the number of linearly polarized pulses in the train, we disentangle the contribution of each pulse to the observed PECD, revealing that the underlying physical mechanism relies on the first linear pulse creating a chiral hole wave packet, which evolves within the chiral molecular ensemble on attosecond timescales, and is later photoionized (probed) by subsequent pulses. Further analysis of the electron-density dynamics directly confirms this mechanism, which is distinct from the standard helical-light PECD mechanism. Our work therefore paves way to novel realizations of PECD and its application for controlled chiral quantum wave packets.

Let us begin by describing our numerical approach. We employed two complementary approaches. The first relies on model single-active-electron simulations. Here we 
solve the  time-dependent Schr\"{o}dinger equation (TDSE) with one active orbital interacting with a train of intense laser pulses (which will be described below). The single electron Hamiltonian and orbital are obtained from ground state density functional theory (DFT)~\cite{engel2011density}, where the highest occupied Kohn-Sham (KS) molecular orbital (HOMO) of CBrClFH is taken as the active orbital since employed laser fields will be in the tunneling regime. We obtain the ground state (GS) of the molecule using the local density approximation. Effectively, this approach is equivalent to performing a TDDFT~\cite{marques2004time,gross1990time} simulation with the first 12 orbitals frozen in time (assuming substantial dynamics can only arise in the HOMO), and where interactions between electrons are neglected (the independent particle approximation). The interaction of the molecule with the driving laser field, $\textbf{E}(t)$, is described in the dipole approximation and length gauge by the TDSE:

\begin{equation}
i\frac{\partial}{\partial t} |\varphi_n(t)\rangle = \left(-\frac{1}{2}\nabla^2 + v(\mathbf{r}) - \mathbf{E}_\Omega(t) \cdot \mathbf{r}\right) |\varphi_n(t)\rangle
\label{eq:EOM}
\end{equation}
where $|\varphi_n(t)\rangle$ is the n-th TD orbital (here $n=13$), initially at t=0 set to be $|\varphi_{\text{HOMO}}\rangle$, $v(\mathbf{r})$ is the effective potential taken to be the KS potential.
By solving the EOM, we obtain the time-dependent wave function, from which the PES is calculated using the t-surff method~\cite{Tao2012,Scrinzi2012} implemented in Octopus\cite{Wopperer2017,Sato2018,DeGiovannini2017}. PECD spectra for each enantiomer are obtained by averaging the driving field $E_{\Omega}(t)$ orientations along the molecular solid angle $\Omega$ using three Euler angles. The equivalent of velocity map imaging (VMI) spectra are calculated along the three main planes (xy, xz, yz) after integrating photoemission in the transverse axis. All additional technical details on the numerical procedures are delegated to the Supplementary Material (SM).

The laser field employed is composed of linearly-polarized pulses, each pulse is formed in the shape of eq.~\ref{eq:E(t)}:
\begin{equation}
E(t,\phi) = E_{0}cos(\omega t+\phi) sin^2\left(\frac{\pi t}{T}\right)
\label{eq:E(t)}
\end{equation}
where, $E_{0}$ is the laser amplitude corresponds to the maximal laser intensity $I_{0}$ in [$W/cm^2$], $\omega$ is the fundamental frequency, $\phi$ is the carrier-envelope phase (CEP) and $T$ is the duration of the pulse. The total laser field consists of a 3-pulse train with varying polarization angles ($\theta$) and time delay ($\Delta t_{1,2}$) (see Figures~\ref{fig:laser} (b) and (c) or full expression in SM)\cite{Neufeld2017njp}. As can be seen in Figure~\ref{fig:laser} (c), the laser is coplanar in polarization-space with identical pulses, i.e. leading to OC = 0 in eq. 1 from ~\citen{tang2010optical}, meaning it is a non-helical light source. 

A second complementary scheme that we employ is solving the full TDDFT KS equations in real-time using Octopus code~\cite{marques2003octopus, andrade2013real, andrade2015realspace}, which is almost identical to the TDSE SAE scheme above, but including contributions from all orbitals in the molecule and allowing electrons to interact with each other (see details in SM, and in ref. \cite{neufeld2021strong}). For simplicity, the majority of the results in the paper are analyzed with the TDSE SAE approach, and we only employ full TDDFT costly simulations for validation.


\begin{figure*}[t]
    \centering
    \includegraphics[width=1.0\linewidth]{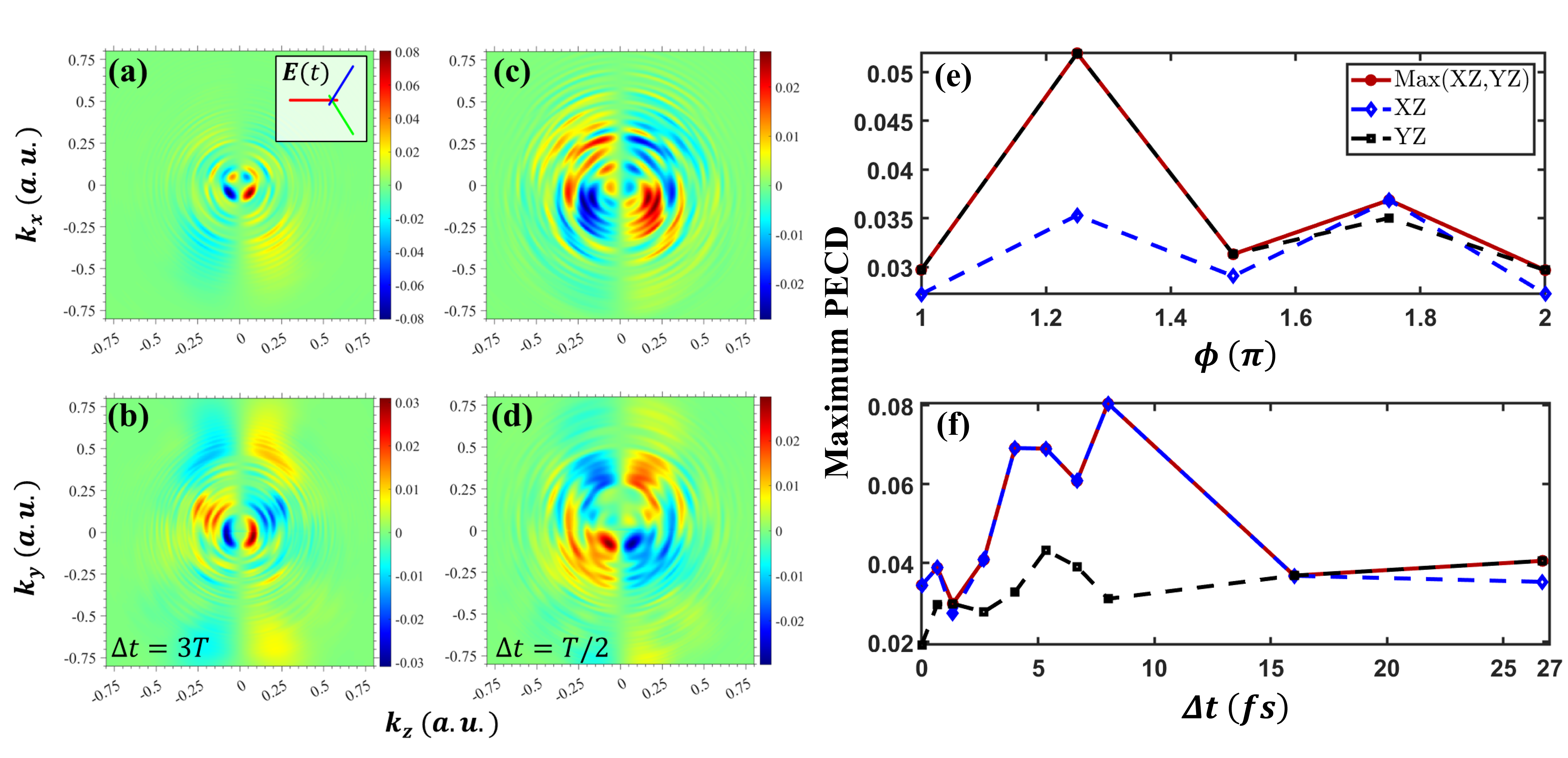}
    \caption{{\bf PECD generated in the SAE model by three-pulse train} with parameters similar to those in Fig.~\ref{fig:laser} (a,b) xz and yz planes PECD for $\Delta t$ = $3T$ respectively. (c,d) xz and yz planes PECD for $\Delta t$ = $T/2$ respectively. Inset in (a) shows the employed laser field lissajous curve. (e) Maximal PECD value vs CEP (for $\Delta t=T/2$). (f) Maximal PECD value vs symmetric delay between pulses (for CEP = $\pi$). }\label{figs:PECD_3pulses}
\end{figure*}

Figure~\ref{figs:PECD_3pulses} (a)-(d) presents examples of the PECD spectra generated in our model framework. Remarkably, we obtain non-zero PECD signal in both xz and yz planes even though the driving laser train is non-helical (for further xy plane PECD see SM). PECD signals are noted to be on similar orders of magnitude to those arising from standard CPL~\cite{beaulieu2017attosecond,PhysRevLett.86.1187} or even bi-chromatic lasers~\cite{neufeld2021strong, neufeld2025symmetries} in similar conditions, reaching ~5\%. 
Our next key goal is to understand the origin of these non-zero chiral signals and the physical mechanisms that drive them. Given the richness of the PECD spectral data, we reduce the complexity of this analysis by focusing on the dependence of the absolute maximum of the chiral signals in each plane on the laser parameters, in hopes of isolating the dominating degrees of freedom.

Figures~\ref{figs:PECD_3pulses} (e) and (f) present the PECD maximum signal dependency on CEP of each pulse in the train, and the delay between pulses in the train, respectively. The results in Figure~\ref{figs:PECD_3pulses} (f) reveal that the maximum chiral signal is extremely sensitive to symmetric delay between pulses ($\Delta t_{1}=\Delta t_{2} = \Delta t$). 
This results in a maximum $\sim$8\% chiral signal when the delay is chosen at $3T$ (Fig.~\ref{figs:PECD_3pulses} (a)), compared to a minimal value of $\sim$2\% for no delay. In contrast, the PECD has weak dependence on the CEP presenting only $\sim$3-5\% modulation in the maximum chiral signal (Figure~\ref{figs:PECD_3pulses} (e)). The dependence on fundamental wavelength (see SM) exhibits $\sim$4.5\% peak at $700 nm$ wavelength and significant chiral signal reduction at longer wavelengths, while more minor variation ($\sim$2.5-3.5\%) in the maximal PECD signal are observed with asymmetric inter-pulse delays. From this analysis we conclude that the intra-pulse parameters in the pulse train such as wavelength and CEP result in expected behavior due to changes to total ionization yield, while the inter-pulse parameters in the pulse train dominantly impact chiral signals, which is not expected from the standard PECD mechanism driven by helical light. The strong delay-dependence suggests attosecond timescale oscillations in the PECD signal, which hints at a coherent hole wave-packet evolving within the chiral molecular ensemble. Thus, the physical mechanism of PECD from non-helical lasers seems to be different from that of the standard CPL case, which relies on the helicity of the laser source. We note that even though the PECD does not vanish in our simulations for longer delay (or longer pulse durations, present in SM), as one might expect, the maximum signal still reduces significantly compared to its maximum. This is a consequence of the lack of dissipation channels in our model, which means the chiral hole wave packet evolves continuously without relaxing, but simply delocalizes over the molecule in longer timescales.

\begin{figure}[H]
    \centering
    \includegraphics[width=1.0\linewidth]{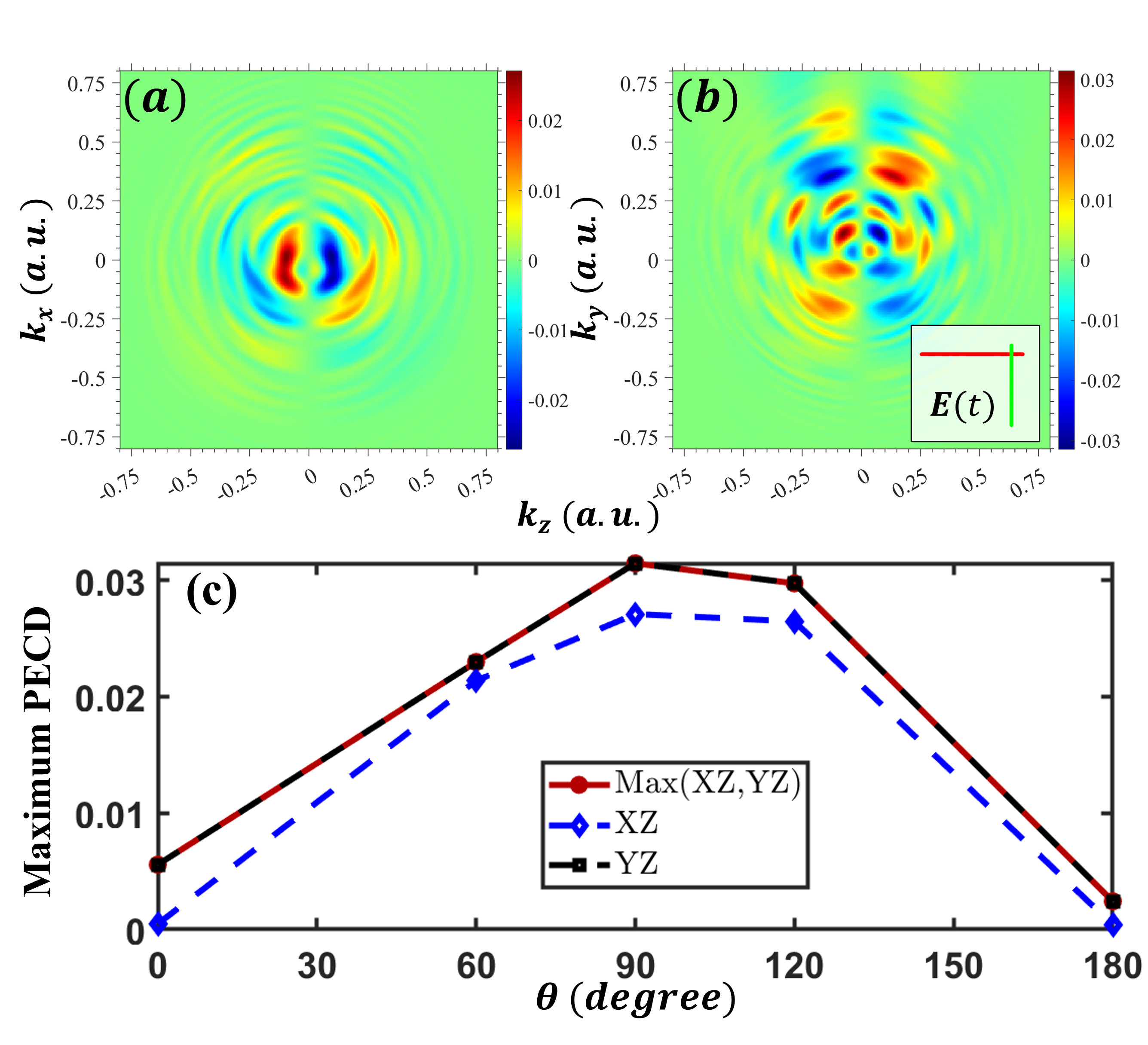}
    \caption{{\bf PECD calculated in the SAE model driven by two-pulse train.} (a) xy, and (b) yz, PECD plane chiral signals, in similar laser field parameters to those in Fig~\ref{fig:laser}, except $\theta$ = $90^o$. Inset in (b) shows the employed laser field lissajous curve. (c) Maximum PECD value vs relative angle between pulse polarization ($\theta$).\label{figs:PECD_2pulses}}
\end{figure}

To further investigate the underlying PECD mechanism and validate that it arises from a cohernet chiral hole wave packet, we reduced the number of pulses in the train to two. The logic here is as follows: one linearly-polarized pulse alone is not able to generate PECD because of the fundamental symmetry of the laser-matter system~\cite{neufeld2025symmetries}; ergo, if a chiral signal is generated in the two-pulse train as well, it must mean it is generated solely by the second pulse.  Figures~\ref{figs:PECD_2pulses} (a) and (b) display PECD obtained by two-pulse trains, exhibiting a similar chiral signal scale and qualitative behavior. Furthermore, by varying the polarization angle ($\theta$) between the two pulses in the train (the inter-pulse relative polarization), we showed that the second pulse must have a certain angle to break the symmetry of the system to generate a chiral signal. Otherwise, we obtain numerically-zero PECD for parallel train (90 or 180 degree angle) (Fig.~\ref{figs:PECD_2pulses} (c)). The above allows us to disentangle the physical mechanism to a two-step process: (i) The first linearly-polarized pulse interacts with the molecule and generates a chiral state in the hole density. Note that this occurs in the averaged molecular ensemble, as well as in each individual orientation of chiral molecule. This pulse therefore acts purely as a pump for chiral dynamics, since it cannot probe chirality due to the fundamental symmetry of the system. (ii) The non-parallel second pulse (or more) interacts with the chiral hole state to produce the PECD from further photo-ionization, acting as probes for the molecular chirality, as well as that of the wave packet. \\

\begin{figure}[h]
    \centering
    \includegraphics[width=1.0\linewidth]{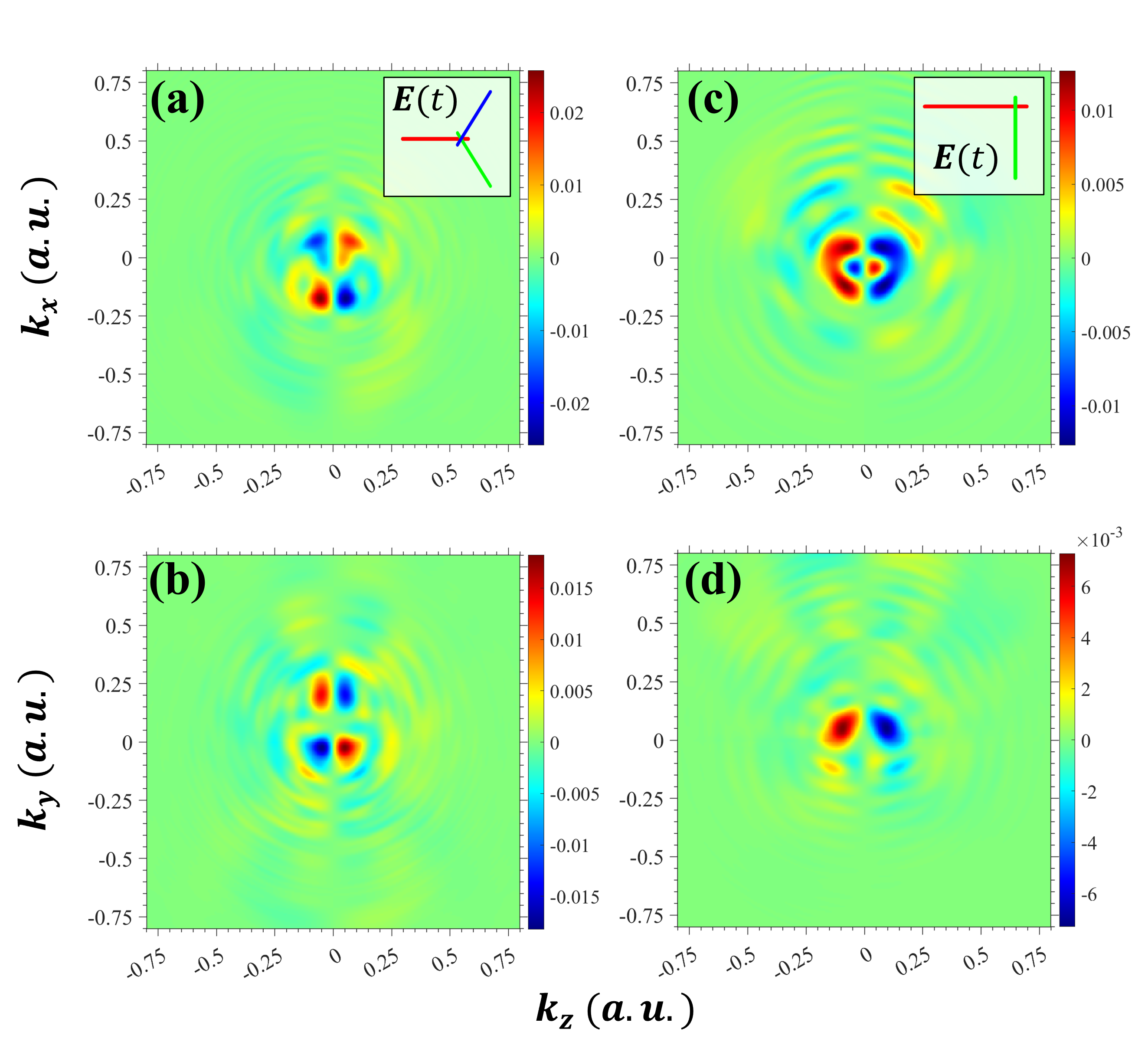}
    \caption{{\bf PECD calculated within full-TDDFT theory for 3- and 2-pulse trains} with similar parameters to Fig~\ref{fig:laser}. (a) and (b) are xz and yz plane PECD for 3 pulses train respectively; and (c) and (d) are xz and yz plane PECD for 2 pulses train respectively. Insets shows the employed laser fields.}\label{figs:PECD_TDDFT}
\end{figure}

To complement this conclusion with direct evidence, we studied the electronic time-dependent density $\rho(\mathbf{r},t)$) dynamic of each enantiomer after introducing the first linear pulse. The density was extracted directly for each molecular orientation, and further orientation averaged. We analyze the difference between the density differences of the enantiomers, i.e. $\Delta\rho _{R}(\mathbf{r},t) - \Delta\rho _{S}(\mathbf{r},t)$, where $\Delta\rho(\mathbf{r},t) = \rho(\mathbf{r},t) - \rho(\mathbf{r},0)$ is the chiral hole density generated after light-matter interactions with the first linear pulse. This data is presented in Fig.~\ref{figs:dynamics} in the Appendix, and shows clear chiral hole evolution in the orientation-averaged ensemble in the sense that the hole wave packet evolves on attosecond timescales in all three main planes. The 3D evolution is shown to be truly chiral since it does not display any mirror symmetries~\cite{kelvin1894molecular,Ayuso2019,Neufeld2020a}. This direct analysis therefore proves our hypothesis for the physical mechanism of PECD from non-helical light. 

Lastly, to verify the validity of our predictions and the model approach, we calculated PECD spectra generated by three- and two-linearly-polarized pulses pulse trains while employing full TDDFT simulations (Figure~\ref{figs:PECD_TDDFT}). We emphasize that these simulations do not neglect deeper molecular orbitals or electron-electron interactions. The results show similar scale PECD, and qualitatively validate the SAE model analysis.


In summary, we have theoretically demonstrated that PECD can arise from a light source with vanishing optical chirality that is non-helical. This result is demonstrated with model calculations, and validated through state-of-the-art \textit{ab-initio} theory. By analyzing the effect of the driving laser source parameters and number of pulses in the pulse train, we showed that the underlying physical mechanism for this type of PECD involves the evolution of a chiral hole-wave packet in the orientation-averaged molecular ensemble\cite{Eckart2018,Wanie2024}, which can be generated by just one linearly-polarized laser pulse. This chiral hole evolution is then probed by subsequent pulses, even in a non-helical set-up, making the physical mechanism distinct from that in standard PECD driven by helical light. We further showed that through this technique chiral signals can straightforwardly be tuned over a wide dynamical range (from $\sim$2-8\%) by controlling attosecond delays between the pump and probe pulses, which follows from the attosecond modulation of the chiral hole wave packet. We expect for PECD from non-helical light to open new avenues for tuning attosecond chiral electron dynamics, and exploring new PECD mechanisms, e.g. by employing bichromatic laser driving to induce hole dynamics and non-helical PECD instead. Another interesting prospect is to probe the chiral hole-wave packet dynamics via pump-probe high-harmonic generation, as well as distinguishing between bound-wave packet motion from that occurring in the cation. Looking ahead, our results also raise questions regarding the connection between chiral signal generation and the non-instantaneous optical chirality\cite{neufeld2018optical} of the non-helical laser. Such insights should also be valuable for the study of magnetism\cite{yang2021chiral,Evers2022,Neufeld2023b}, and chiral topological systems\cite{Lv2021,Bharti2024,Yan2017}.
\\

\noindent {\bf Acknowledgment}\\
 ON gratefully acknowledges the scientific support of Prof. Dr. Angel Rubio, and support from the Young Faculty Award from the National Quantum Science and Technology program of Israel's Council of Higher Education Planning and Budgeting Committee.\\

\appendix
\section{Chiral hole-wavepacket dynamics}
\begin{figure*}[b]
    \centering
    \includegraphics[width=\linewidth]{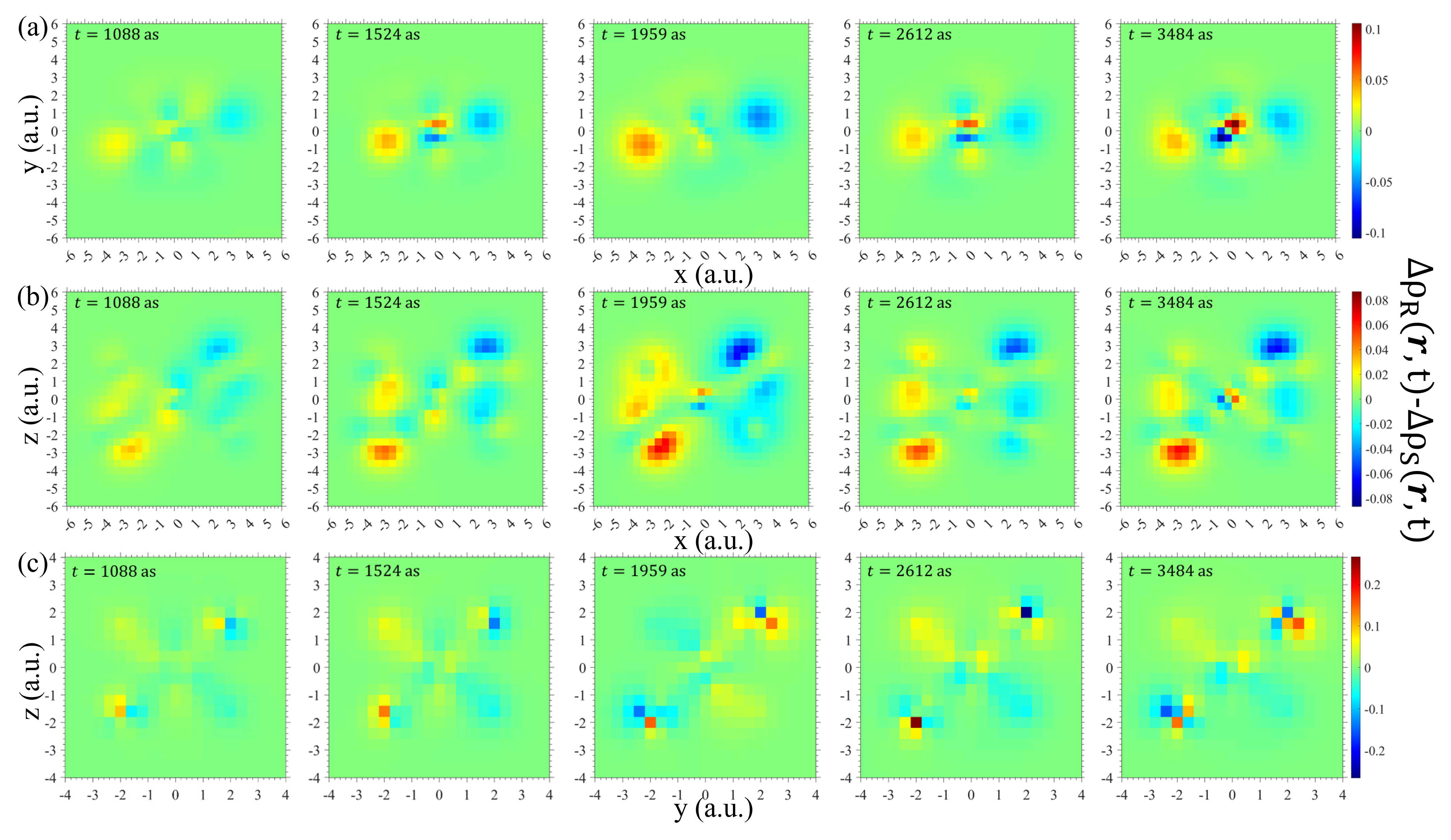}
    \caption{{\bf Orientation-averaged chiral hole state density difference ($\Delta\rho _{R}(\mathbf{r},t) - \Delta\rho _{S}(\mathbf{r},t)$) dynamics in (a) xy molecular plane, (b) xz molecular plane, and (c) yz molecular plane.} Generated in the SAE model by one linear pulse with intensity of $5\cdot10^{13} [W/cm^2]$, $\omega$ corresponds to a wavelength of $800nm$ and one optical cycle of the fundamental duration $T$. The total simulation duration is $3T$ with $dt = 218$ $[as]$. The data demonstrate clear chiral hole evolution following the first linear pulse that acts as a pump (inducing chiral dynamics in 3D).}\label{figs:dynamics}
\end{figure*}

We computed the orientation-averaged, time-dependent electronic density $\rho(\mathbf{r},t)$ for both enantiomers after the first linearly polarized pulse, using the same orientation averaging procedure as in the PES calculations. Figure~\ref{figs:dynamics} displays the difference between their induced density changes, $\Delta\rho_R(\mathbf{r},t) - \Delta\rho_S(\mathbf{r},t)$, where $\Delta\rho(\mathbf{r},t)=\rho(\mathbf{r},t)-\rho(\mathbf{r},0)$. These dynamics reveal a clear chiral hole wave-packet evolution. The attosecond motion persists across all principal planes and lacks any mirror symmetry, consistent with genuinely 3D chiral behavior~\cite{kelvin1894molecular,Ayuso2019,Neufeld2020a}.

\clearpage
\newpage
\bibliography{ref}

\end{document}


\preprint{APS/123-QED}

\title{Supplementary Material of \\Photoelectron chiral dichroism from lasers without helicity via chiral hole wave-packet} 

\author{Gal Bouskila}\email{galboski@campus.technion.ac.il}
\affiliation{Schulich Faculty of Chemistry, Technion – Israel Institute of Technology, 32000 Haifa, Israel}
\author{Avner Fleischer}
\affiliation{
 Raymond and Beverly Sackler Faculty of Exact Sciences, School of Chemistry, Tel Aviv University, Tel-Aviv, Israel}
\author{Ofer Neufeld}\email{ofern@technion.ac.il}
\affiliation{Schulich Faculty of Chemistry, Technion – Israel Institute of Technology, 32000 Haifa, Israel}

\date{\today}

\maketitle
We present here further numerical details of the simulations as well as additional results complementary to those in the main text. All simulations were performed using Octopus code~\cite{castro2006octopus, andrade2015realspace, TancogneDejean2020}. The equations of motion were discretized on a Cartesian grid with spherical boundaries of radius 40 Bohr, where the molecule center of mass was centered at the origin. The molecular geometry was taken at the experimental configuration. The grid spacing was converged to 0.4 Bohr. For time-dependent calculations, we employed a time step $\Delta t=0.18$ a.u., and added an imaginary absorbing potential of width 10 Bohr. The initial state was taken to be the system’s ground-state. The full photoelectron spectrum from each driving laser orientation was calculated using the t-surff method~\cite{Tao2012,Scrinzi2012}, implemented within Octopus code\cite{Wopperer2017,Sato2018,DeGiovannini2017}. The flux was calculated over the spherical surface with at r=30 Bohr, and integration was performed with a maximal angular momentum index for spherical harmonics of 80, angular grids were spanned with a 1 degree spacing, k-grids were spanned with a spacing of $\Delta k=0.01$ a.u. and up to a maximal energy of $25\cdot\omega$ eV. The orientation averaged PES was calculated by trapezoidal integration, similar to the method in ref~\citen{neufeld2021strong}. Photoelectron chiral dichroism spectra were obtained directly by subtracting the photoemission calculated from the mirror image enantiomers. The equivalent of velocity map imaging (VMI) spectra are calculated along the three main planes (xy, xz, yz) after integrating photoemission in the transverse axis. For example, PECD in the xy plane is obtained by integrating the component $k_{z}$ out, and similarly for other axes.

\section{Time dependent density functional theory}
Time dependent density functional theory (TDDFT) calculations were performed using Octopus code~\cite{marques2003time}. The interaction of the molecule with the driving laser field, $\mathbf{E}(t)$, is described in the dipole approximation and length gauge by the Kohn-Sham (KS) equations of motion:

\begin{equation}
    i\partial_t |\varphi_n^{KS}(t)\rangle = \left(-\frac{1}{2}\nabla^2 + v_{KS}(\mathbf{r},t) - \mathbf{E}(t) \cdot \mathbf{r}\right) |\varphi_n^{KS}(t)\rangle
\end{equation}

where $|\varphi_n^{KS}(t)\rangle$ is the n'th time-dependent KS orbital, $v_{KS}(\mathbf{r},t)$ is the time-dependent KS potential:

\begin{equation}
    v_{KS}(\mathbf{r},t) = V_{ion} + \int d^3r' \frac{\rho(\mathbf{r}',t)}{|\mathbf{r} - \mathbf{r}'|} + v_{XC}[\rho(\mathbf{r},t)]
\end{equation}

Here, $V_{ion}$ is an electrostatic interaction of electrons with nuclei and other core electrons using pseudopotentials~\cite{hartwigsen1998relativistic}, and assuming static ions. $\rho(\mathbf{r},t)$ is the electronic density and $v_{XC}$ is the exchange correlation (XC) potential taken to be the Perdew and Zunger self-interaction-corrected adiabatic local-density approximation (LDA)~\cite{legrand2002comparison}.

\section{Full expression of the laser field}
The total laser field employed consists of a train of 3 pulses (except for the two-pulse train discussed in the main text), each pulse is formed in the shape of eq. (3) in the main text, starting with an x-polarized pulse. After the end of the first pulse and a short delay ($\Delta t_1$), a second $\theta = 120^\circ$ rotated pulse in the polarization space. Finally, the third pulse is added after the end of the second pulse and a short delay ($\Delta t_2$), which is also rotated $\theta = 120^\circ$ in the polarization space. The relation of the 3 pulses can be written as:

\begin{equation}
\begin{split}
    E_{Tot}(t)   & = E_{1}(t,\phi_1)(\hat{x},0,0)(1-u_{T^f_1}(t))\\
    & + E_{2}(t,\phi_2)(\hat{x}cos(\theta),\hat{y}sin(\theta),0)(u_{T^i_2}(t)(1-u_{T^f_2}(t)))\\
    & + E_{3}(t,\phi_3)(\hat{x}cos(2\theta),\hat{y}sin(2\theta),0)u_{T^i_3}(t)
\end{split}
\label{eq:E(t)_tot}
\end{equation}

where $\phi_2 = -\omega(\Delta t_1+T)+\phi_1$ and $\phi_3 = -\omega(\Delta t_1+ \Delta t_2+2T)+\phi_1$ are phase corrections to ensure phase-coherent of the pulses with respect to the first pulse and $\omega$ is the frequency. $u_a(t)$ is the step-function equal to 1 for $t > a$ and 0 otherwise; $T^i_{no.}$ and $T^f_{no.}$ are the start and end time of pulse no. (=1,2,3), respectively, and are equal to $T^i_1=0$, $T^f_1=T$, $T^f_2=T+\Delta t_1$,$T^f_2=2T+\Delta t_1$, $T^i_3=2T+\Delta t_1+\Delta t_2$ and $T^f_3=3T+\Delta t_1+\Delta t_2$. 

\section{Additional results}
\begin{figure}[htp]
    \centering
    \includegraphics[width=1\linewidth]{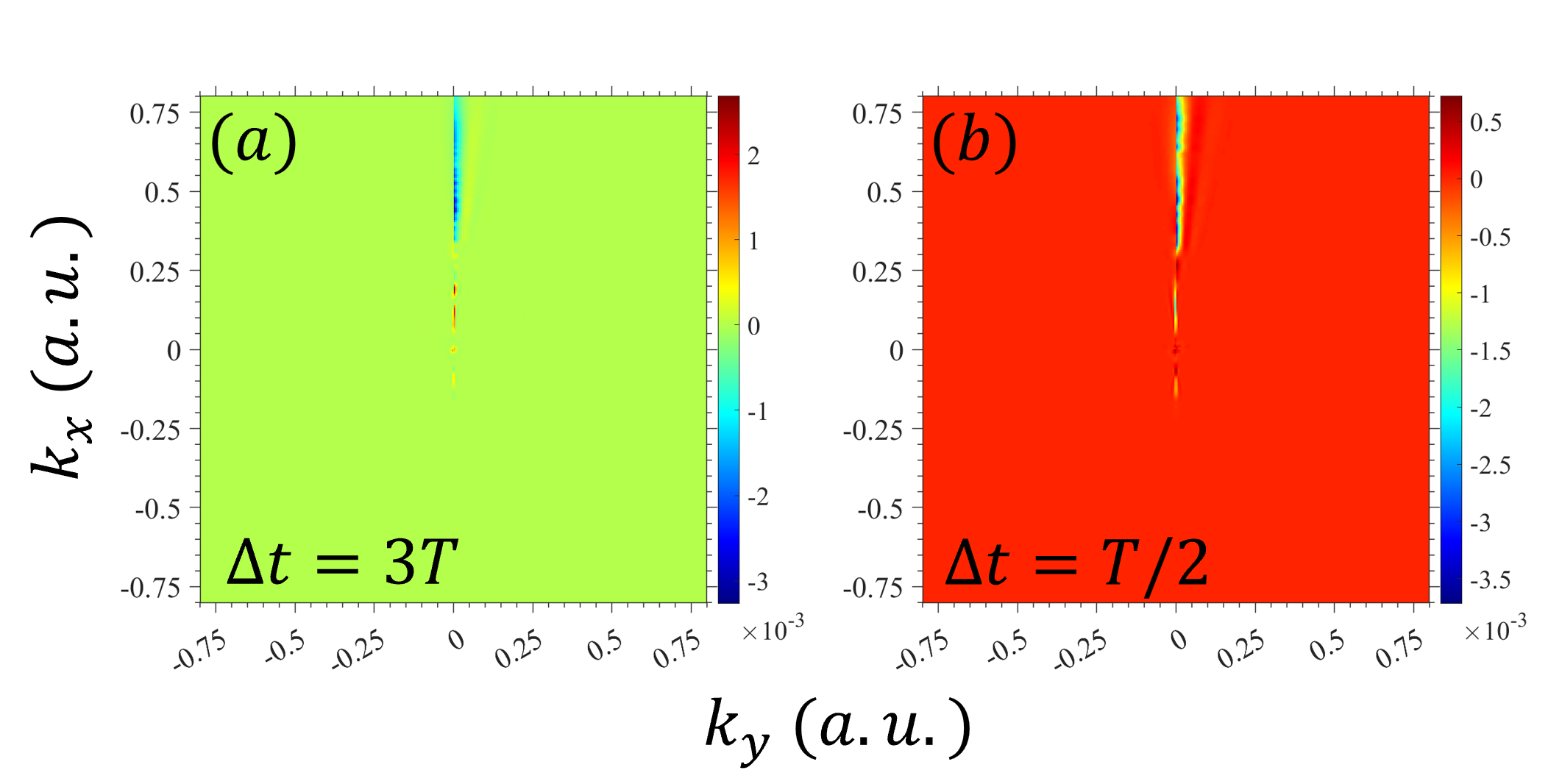}
    \caption{{\bf xy plane PECD generated in the SAE model by three-pulses train} with intensity of $5\cdot10^{13} [W/cm^2]$, $\theta = 120^o$, CEP = $\pi$, $\omega$ corresponds to fundamental wavelength of $800nm$, one optical cycle of fundamental duration T. (a) for symmetric delay = $3T$  and (b) for symmetric delay = $T/2$. Those graphs are relate to Fig. 2 in the main letter}\label{SM_figs:pecd_sae_xy}
\end{figure}

Figure~\ref{SM_figs:pecd_sae_xy} presents examples of the xy-planes PECD obtained in our framework. As can be expected, due to the fundamental symmetries~\cite{neufeld2025symmetries} of the system, the results are numerically zero, yielding a sanity check for the numerical scheme.

\begin{figure}[htp]
    \centering
    \includegraphics[width=0.6\linewidth]{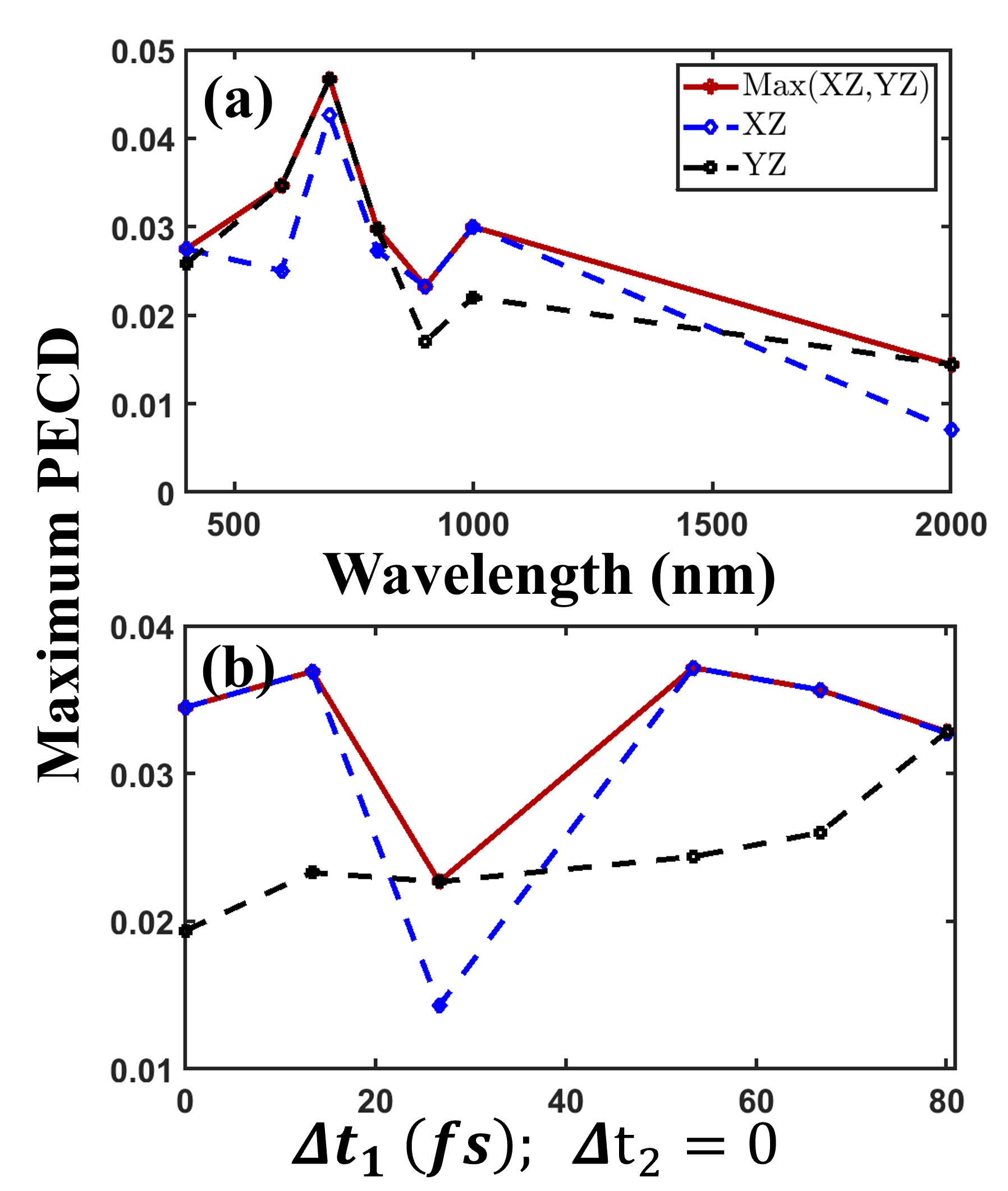}
    \caption{{\bf Maximal PECD value vs (a) wavelength ($\Delta t = T/2$) and (b) asymmetric delay ($\Delta t_2 = 0$, $\omega$ corresponds to wavelength of $800nm$) in the SAE model by three-pulses train} with intensity of $5\cdot10^{13} [W/cm^2]$, $\theta = 120^o$, CEP = $\pi$ and one optical cycle of the fundamental duration T.}\label{SM_figs:pecd_wl_delay}
\end{figure}

Figures~\ref{SM_figs:pecd_wl_delay} (a) and (b) present the PECD maximum signal dependency on wavelength of each pulse in the train, and the asymmetric delay (changes in $\Delta t_1$, while $\Delta t_2 = 0$) between pulses in the train, respectively. The results in Figure~\ref{SM_figs:pecd_wl_delay} reveal weak dependency of the PECD on laser parameter. Figure~\ref{SM_figs:pecd_wl_delay} (a) exhibits $\sim$4.5\% peak at $700 nm$ wavelength and significant chiral signal reduction at longer wavelengths, while Fig.~\ref{SM_figs:pecd_wl_delay} (b) show more minor variation ($\sim$2.5-3.5\%) in the maximal PECD signal are observed with asymmetric inter-pulse delays. 

\begin{figure}[htp]
    \centering
    \includegraphics[width=0.6\linewidth]{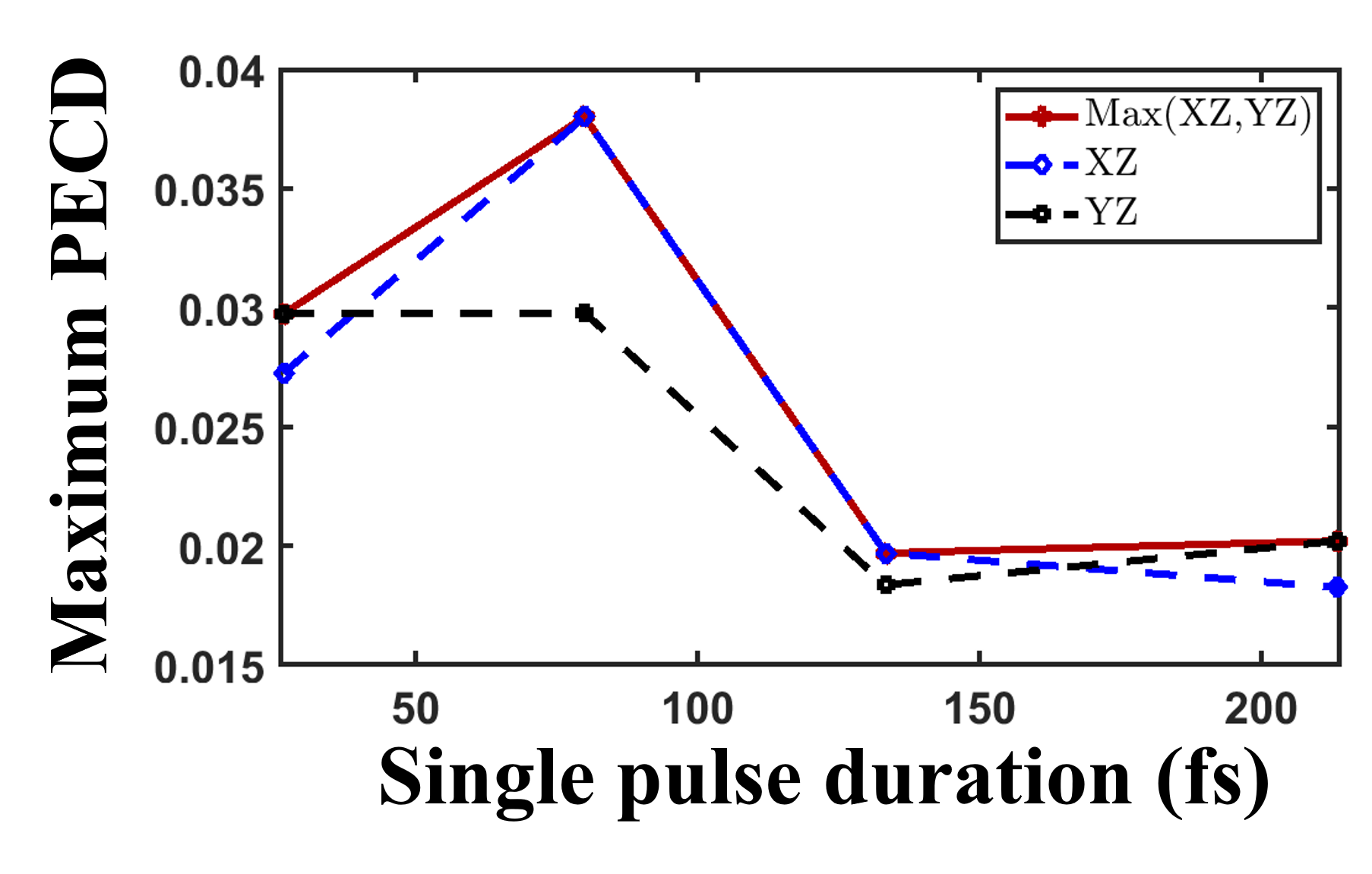}
    \caption{{\bf Maximal PECD value vs single pulse duration in the SAE model by three-pulses train} with intensity of $5\cdot10^{13} [W/cm^2]$, $\theta = 120^o$, CEP = $\pi$ and $\Delta t = T/2$.}\label{SM_figs:pecd_cycl}
\end{figure}

Fig.~\ref{SM_figs:pecd_cycl} presents the maximal PECD value dependency on the individual pulse duration, which corresponds to the total pulse-train duration. We note that even though the PECD does not vanish in our simulations for longer duration and delay, the maximum signal still considerably reduces compared to its maximum. As mentioned in the main text, the non-vanishing signal is due to the lack of dissipation channels in our model, meaning the chiral hole wave packet evolves continuously without relaxing, but delocalizes over the molecule in longer timescales.

\clearpage
\bibliography{ref}